\input harvmac
\input amssym
%\draftmode
\baselineskip 13pt

\def\p{\partial}

\def\half{{1\over 2}}
\def\rar{\rightarrow}

\def\vphi{\varphi}
\def\vs{\vskip .1 in}
\def\a{\alpha}
\def\b{\beta}

\def\p{\partial}
\def\bul{$\bullet$~}

\def\IR{\relax{\rm I\kern-.18em R}}
\def\BR{\IR}

\def\p{\partial}

\def\half{{1\over 2}}
\def\rar{\rightarrow}

\def\vphi{\varphi}
\def\vs{\vskip .1 in}
\def\a{\alpha}
\def\b{\beta}

\def\p{\partial}

\def\rar{\rightarrow}

\def\xb{{\bf x}}

%%%%%%%%%%%%%%%%%%%%%%%%%%% REFS:

%\TaylorTG
\lref\TaylorTG{
  M.~Taylor,
  ``Non-relativistic holography,''
[arXiv:0812.0530 [hep-th]].
%%CITATION = arXiv:0812.0530%%
}

\lref\aprile{
  F.~Aprile and J.~G.~Russo,
  ``Models of Holographic superconductivity,''
Phys.\ Rev.\ D {\bf 81}, 026009 (2010).
[arXiv:0912.0480 [hep-th]].
%%CITATION = arXiv:0912.0480%%
}

%\DeyRS
\lref\DeyRS{
  P.~Dey and S.~Roy,
  ``Intersecting D-branes and Lifshitz-like space-time,''\vskip .001 in
[arXiv:1204.4858 [hep-th]].
%%CITATION = arXiv:1204.4858%%
}

%\KimNB
\lref\KimNB{
  B.~S.~Kim,
  ``Schr\'odinger Holography with and without Hyperscaling Violation,''
[arXiv:1202.6062 [hep-th]].
%%CITATION = arXiv:1202.6062%%
}

%\NarayanHK
\lref\NarayanHK{
  K.~Narayan,
  ``On Lifshitz scaling and hyperscaling violation in string theory,''
[arXiv:1202.5935 [hep-th]].
%%CITATION = arXiv:1202.5935%%
}

%\SinghZS
\lref\SinghZS{
  H.~Singh,
  ``Special limits and non-relativistic solutions,''
JHEP {\bf 1012}, 061 (2010).
[arXiv:1009.0651 [hep-th]].
%%CITATION = arXiv:1009.0651%%
}

%\GouterauxCE
\lref\GouterauxCE{
  B.~Gouteraux and E.~Kiritsis,
  ``Generalized Holographic Quantum Criticality at Finite Density,''
JHEP {\bf 1112}, 036 (2011).
[arXiv:1107.2116 [hep-th]].
%%CITATION = arXiv:1107.2116%%
}

%\BobevRV
\lref\BobevRV{
  N.~Bobev, A.~Kundu, K.~Pilch and N.~P.~Warner,
  ``Minimal Holographic Superconductors from Maximal Supergravity,''
JHEP {\bf 1203}, 064 (2012).
[arXiv:1110.3454 [hep-th]].
%%CITATION = arXiv:1110.3454%%
}

%\CharmousisZZ
\lref\CharmousisZZ{
  C.~Charmousis, B.~Gouteraux, B.~S.~Kim, E.~Kiritsis and R.~Meyer,
  ``Effective Holographic Theories for low-temperature condensed matter systems,''
JHEP {\bf 1011}, 151 (2010).
[arXiv:1005.4690 [hep-th]].
%%CITATION = arXiv:1005.4690%%
}

%\DongSE
\lref\DongSE{
  X.~Dong, S.~Harrison, S.~Kachru, G.~Torroba and H.~Wang,
  ``Aspects of holography for theories with hyperscaling violation,''
[arXiv:1201.1905 [hep-th]].
%%CITATION = arXiv:1201.1905%%
}

%\HuijseEF
\lref\HuijseEF{
  L.~Huijse, S.~Sachdev and B.~Swingle,
  ``Hidden Fermi surfaces in compressible states of gauge-gravity duality,''
Phys.\ Rev.\ B {\bf 85}, 035121 (2012).
[arXiv:1112.0573 [cond-mat.str-el]].
%%CITATION = arXiv:1112.0573%%
}

%\GubserQT
\lref\GubserQT{
  S.~S.~Gubser and F.~D.~Rocha,
  ``Peculiar properties of a charged dilatonic black hole in AdS$_5$,''
Phys.\ Rev.\ D {\bf 81}, 046001 (2010).
[arXiv:0911.2898 [hep-th]].
%%CITATION = arXiv:0911.2898%%
}

%\BalasubramanianUW
\lref\BalasubramanianUW{
  K.~Balasubramanian and J.~McGreevy,
  ``The Particle number in Galilean holography,''
JHEP {\bf 1101}, 137 (2011).
[arXiv:1007.2184 [hep-th]].
%%CITATION = arXiv:1007.2184%%
}

%\ShaghoulianAA
\lref\ShaghoulianAA{
  E.~Shaghoulian,
  ``Holographic Entanglement Entropy and Fermi Surfaces,''\vskip .001 in 

[arXiv:1112.2702 [hep-th]].
%%CITATION = arXiv:1112.2702%%
}

%\PerlmutterQU
\lref\PerlmutterQU{
  E.~Perlmutter,
  ``Domain Wall Holography for Finite Temperature Scaling Solutions,''
JHEP {\bf 1102}, 013 (2011).
[arXiv:1006.2124 [hep-th]].
%%CITATION = arXiv:1006.2124%%
}

%\JeongAA
\lref\JeongAA{
  J.~Jeong, H.~-C.~Kim, S.~Lee, E.~O Colgain and H.~Yavartanoo,
  ``Schrodinger invariant solutions of M-theory with Enhanced Supersymmetry,''
JHEP {\bf 1003}, 034 (2010).
[arXiv:0911.5281 [hep-th]].
%%CITATION = arXiv:0911.5281%%
}

%\KrausPF
\lref\KrausPF{
  P.~Kraus and E.~Perlmutter,
  ``Universality and exactness of Schrodinger geometries in string and M-theory,''
JHEP {\bf 1105}, 045 (2011).
[arXiv:1102.1727 [hep-th]].
%%CITATION = arXiv:1102.1727%%
}

%\BobevQX
\lref\BobevQX{
  N.~Bobev and B.~C.~van Rees,
  ``Schrodinger Deformations of $AdS_3 x S^3$,''
JHEP {\bf 1108}, 062 (2011).
[arXiv:1102.2877 [hep-th]].
%%CITATION = arXiv:1102.2877%%
}

%\HartnollWM
\lref\HartnollWM{
  S.~A.~Hartnoll and E.~Shaghoulian,
  ``Spectral weight in holographic scaling geometries,''
[arXiv:1203.4236 [hep-th]].
%%CITATION = arXiv:1203.4236%%
}

%\MazzucatoTR
\lref\MazzucatoTR{
  L.~Mazzucato, Y.~Oz and S.~Theisen,
  ``Non-relativistic Branes,''
JHEP {\bf 0904}, 073 (2009).
[arXiv:0810.3673 [hep-th]].
%%CITATION = arXiv:0810.3673%%
}

%\GouterauxQH
\lref\GouterauxQH{
  B.~Gouteraux, J.~Smolic, M.~Smolic, K.~Skenderis and M.~Taylor,
  ``Holography for Einstein-Maxwell-dilaton theories from generalized dimensional reduction,''
JHEP {\bf 1201}, 089 (2012).
[arXiv:1110.2320 [hep-th]].
%%CITATION = arXiv:1110.2320%%
}

%\CveticPU
\lref\CveticPU{
  M.~Cvetic, J.~T.~Liu, H.~Lu and C.~N.~Pope,
  ``Domain wall supergravities from sphere reduction,''
Nucl.\ Phys.\ B {\bf 560}, 230 (1999).
[hep-th/9905096].
%%CITATION = hep-th/9905096%%
}

%\ElizaldeNB
\lref\ElizaldeNB{
  E.~Elizalde, M.~Lygren and D.~V.~Vassilevich,
  ``Antisymmetric tensor fields on spheres: Functional determinants and nonlocal counterterms,''
J.\ Math.\ Phys.\  {\bf 37}, 3105 (1996).
[hep-th/9602113].
%%CITATION = hep-th/9602113%%
}

%\ItzhakiDD
\lref\ItzhakiDD{
  N.~Itzhaki, J.~M.~Maldacena, J.~Sonnenschein and S.~Yankielowicz,
  ``Supergravity and the large N limit of theories with sixteen supercharges,''
Phys.\ Rev.\ D {\bf 58}, 046004 (1998).
[hep-th/9802042].
%%CITATION = hep-th/9802042%%
}

%\DonosEN
\lref\DonosEN{
  A.~Donos and J.~P.~Gauntlett,
  ``Supersymmetric solutions for non-relativistic holography,''
JHEP {\bf 0903}, 138 (2009).
[arXiv:0901.0818 [hep-th]].
%%CITATION = arXiv:0901.0818%%
}

%\DonosXC
\lref\DonosXC{
  A.~Donos and J.~P.~Gauntlett,
  ``Solutions of type IIB and D=11 supergravity with Schrodinger(z) symmetry,''
JHEP {\bf 0907}, 042 (2009).
[arXiv:0905.1098 [hep-th]].
%%CITATION = arXiv:0905.1098%%
}
%\DonosTU
\lref\DonosTU{
  A.~Donos and J.~P.~Gauntlett,
  ``Lifshitz Solutions of D=10 and D=11 supergravity,''
JHEP {\bf 1012}, 002 (2010).
[arXiv:1008.2062 [hep-th]].
%%CITATION = arXiv:1008.2062%%
}
%\BobevMW
\lref\BobevMW{
  N.~Bobev, A.~Kundu and K.~Pilch,
  ``Supersymmetric IIB Solutions with Schrodinger Symmetry,''
JHEP {\bf 0907}, 107 (2009).
[arXiv:0905.0673 [hep-th]].
%%CITATION = arXiv:0905.0673%%
}
%\KoroteevYP
\lref\KoroteevYP{
  P.~Koroteev and M.~Libanov,
  ``On Existence of Self-Tuning Solutions in Static Braneworlds without Singularities,''
JHEP {\bf 0802}, 104 (2008).
[arXiv:0712.1136 [hep-th]].
%%CITATION = arXiv:0712.1136%%
}

%\SinghUN
\lref\SinghUN{
  H.~Singh,
  ``Lifshitz/Schr\'odinger Dp-branes and dynamical exponents,''\vskip .001 in
[arXiv:1202.6533 [hep-th]].
%%CITATION = arXiv:1202.6533%%
}

%\DeyTG
\lref\DeyTG{
  P.~Dey and S.~Roy,
  ``Lifshitz-like space-time from intersecting branes in string/M theory,''
[arXiv:1203.5381 [hep-th]].
%%CITATION = arXiv:1203.5381%%
}

%\KachruYH
\lref\KachruYH{
  S.~Kachru, X.~Liu and M.~Mulligan,
  ``Gravity Duals of Lifshitz-like Fixed Points,''
Phys.\ Rev.\ D {\bf 78}, 106005 (2008).
[arXiv:0808.1725 [hep-th]].
%%CITATION = arXiv:0808.1725%%
}

%\SonYE
\lref\SonYE{
  D.~T.~Son,
  ``Toward an AdS/cold atoms correspondence: A Geometric realization of the Schrodinger symmetry,''
Phys.\ Rev.\ D {\bf 78}, 046003 (2008).
[arXiv:0804.3972 [hep-th]].
%%CITATION = arXiv:0804.3972%%
}

%\BalasubramanianDM
\lref\BalasubramanianDM{
  K.~Balasubramanian and J.~McGreevy,
  ``Gravity duals for non-relativistic CFTs,''
Phys.\ Rev.\ Lett.\  {\bf 101}, 061601 (2008).
[arXiv:0804.4053 [hep-th]].
%%CITATION = arXiv:0804.4053%%
}

%\GuicaSW
\lref\GuicaSW{
  M.~Guica, K.~Skenderis, M.~Taylor and B.~C.~van Rees,
  ``Holography for Schrodinger backgrounds,''
JHEP {\bf 1102}, 056 (2011).
[arXiv:1008.1991 [hep-th]].
%%CITATION = arXiv:1008.1991%%
}

%\SongSR
\lref\SongSR{
  W.~Song and A.~Strominger,
  ``Warped AdS3/Dipole-CFT Duality,''
[arXiv:1109.0544 [hep-th]].
%%CITATION = arXiv:1109.0544%%
}

%\MaldacenaWH
\lref\MaldacenaWH{
  J.~Maldacena, D.~Martelli and Y.~Tachikawa,
  ``Comments on string theory backgrounds with non-relativistic conformal symmetry,''
JHEP {\bf 0810}, 072 (2008).
[arXiv:0807.1100 [hep-th]].
%%CITATION = arXiv:0807.1100%%
}

%\SachdevDQ
\lref\SachdevDQ{
  S.~Sachdev,
  ``The Quantum phases of matter,''
[arXiv:1203.4565 [hep-th]].
%%CITATION = arXiv:1203.4565%%
}

%\OgawaBZ
\lref\OgawaBZ{
  N.~Ogawa, T.~Takayanagi and T.~Ugajin,
  ``Holographic Fermi Surfaces and Entanglement Entropy,''
JHEP {\bf 1201}, 125 (2012).
[arXiv:1111.1023 [hep-th]].
%%CITATION = arXiv:1111.1023%%
}

%\CassaniSV
\lref\CassaniSV{
  D.~Cassani and A.~F.~Faedo,
  ``Constructing Lifshitz solutions from AdS,''
JHEP {\bf 1105}, 013 (2011).
[arXiv:1102.5344 [hep-th]].
%%CITATION = arXiv:1102.5344%%
}

%\HartnollRS
\lref\HartnollRS{
  S.~A.~Hartnoll and K.~Yoshida,
  ``Families of IIB duals for nonrelativistic CFTs,''
JHEP {\bf 0812}, 071 (2008).
[arXiv:0810.0298 [hep-th]].
%%CITATION = arXiv:0810.0298%%
}

%\LuRHB
\lref\LuRHB{
  H.~Lu, C.~N.~Pope and P.~K.~Townsend,
  ``Domain walls from anti-de Sitter space-time,''
Phys.\ Lett.\ B {\bf 391}, 39 (1997).
[hep-th/9607164].
%%CITATION = hep-th/9607164%%
}

%\LuCS
\lref\LuCS{
  H.~Lu, C.~N.~Pope, E.~Sezgin and K.~S.~Stelle,
  ``Stainless super p-branes,''
Nucl.\ Phys.\ B {\bf 456}, 669 (1995).
[hep-th/9508042].
%%CITATION = hep-th/9508042%%
}

%\LuYN
\lref\LuYN{
  H.~Lu and C.~N.~Pope,
  ``P-brane solitons in maximal supergravities,''
Nucl.\ Phys.\ B {\bf 465}, 127 (1996).
[hep-th/9512012].
%%CITATION = hep-th/9512012%%
}

%\LavrinenkoMP
\lref\LavrinenkoMP{
  I.~V.~Lavrinenko, H.~Lu and C.~N.~Pope,
  ``From topology to generalized dimensional reduction,''
Nucl.\ Phys.\ B {\bf 492}, 278 (1997).
[hep-th/9611134].
%%CITATION = hep-th/9611134%%
}

%\BoonstraMP
\lref\BoonstraMP{
  H.~J.~Boonstra, K.~Skenderis and P.~K.~Townsend,
  ``The domain wall / QFT correspondence,''
JHEP {\bf 9901}, 003 (1999).
[hep-th/9807137].
%%CITATION = hep-th/9807137%%
}

%\LuNT
\lref\LuNT{
  H.~Lu and C.~N.~Pope,
  ``p-brane taxonomy,''
In *Trieste 1996, High energy physics and cosmology* 340-384.
[hep-th/9702086].
%%CITATION = hep-th/9702086%%
}

%\DuffHP
\lref\DuffHP{
  M.~J.~Duff, H.~Lu and C.~N.~Pope,
  ``The Black branes of M theory,''
Phys.\ Lett.\ B {\bf 382}, 73 (1996).
[hep-th/9604052].
%%CITATION = hep-th/9604052%%
}

%\CveticPN
\lref\CveticPN{
  M.~Cvetic, H.~Lu and C.~N.~Pope,
  ``Domain walls with localized gravity and domain wall / QFT correspondence,''
Phys.\ Rev.\ D {\bf 63}, 086004 (2001).
[hep-th/0007209].
%%CITATION = hep-th/0007209%%
}

%\LuHB
\lref\LuHB{
  H.~Lu, C.~N.~Pope, T.~A.~Tran and K.~W.~Xu,
  ``Classification of p-branes, NUTs, waves and intersections,''
Nucl.\ Phys.\ B {\bf 511}, 98 (1998).
[hep-th/9708055].
%%CITATION = hep-th/9708055%%
}

%\ScherkZR
\lref\ScherkZR{
  J.~Scherk and J.~H.~Schwarz,
  ``How to Get Masses from Extra Dimensions,''
Nucl.\ Phys.\ B {\bf 153}, 61 (1979)..
%%CITATION = LPTENS-79-2%%
}

%\GauntlettDN
\lref\GauntlettDN{
  J.~P.~Gauntlett, J.~Sonner and T.~Wiseman,
  ``Holographic superconductivity in M-Theory,''
Phys.\ Rev.\ Lett.\  {\bf 103}, 151601 (2009).
[arXiv:0907.3796 [hep-th]].
%%CITATION = arXiv:0907.3796%%
}

%\HartnollVX
\lref\HartnollVX{
  S.~A.~Hartnoll, C.~P.~Herzog and G.~T.~Horowitz,
  ``Building a Holographic Superconductor,''
Phys.\ Rev.\ Lett.\  {\bf 101}, 031601 (2008).
[arXiv:0803.3295 [hep-th]].
%%CITATION = arXiv:0803.3295%%
}

%\GubserQM
\lref\GubserQM{
  S.~S.~Gubser, C.~P.~Herzog, S.~S.~Pufu and T.~Tesileanu,
  ``Superconductors from Superstrings,''
Phys.\ Rev.\ Lett.\  {\bf 103}, 141601 (2009).
[arXiv:0907.3510 [hep-th]].
%%CITATION = arXiv:0907.3510%%
}

%\DenefTP
\lref\DenefTP{
  F.~Denef and S.~A.~Hartnoll,
  ``Landscape of superconducting membranes,''
Phys.\ Rev.\ D {\bf 79}, 126008 (2009).
[arXiv:0901.1160 [hep-th]].
%%CITATION = arXiv:0901.1160%%
}

%\DonosZF
\lref\DonosZF{
  A.~Donos and J.~P.~Gauntlett,
  ``Schrodinger invariant solutions of type IIB with enhanced supersymmetry,''
JHEP {\bf 0910}, 073 (2009).
[arXiv:0907.1761 [hep-th]].
%%CITATION = arXiv:0907.1761%%
}

%%%%%%%%%%%%%%%%%%%%%%%%%%%%

\Title{\vbox{\baselineskip14pt
}} {\vbox{\centerline {Hyperscaling violation from supergravity}}}
\centerline{Eric Perlmutter\foot{perl@physics.ucla.edu}}
\bigskip
\centerline{\it{Department of Physics and Astronomy}}
\centerline{${}$\it{University of California, Los Angeles, CA 90095, USA}}

\baselineskip14pt

\vskip .3in

\centerline{\bf Abstract}
\vskip.2cm
In recent applications of AdS/CFT to condensed matter physics, a metric that transforms covariantly under dilatation has been argued to signal hyperscaling violation in a dual quantum field theory. We contextualize and introduce large, in some cases infinite, families of supergravity solutions with this property, focusing on scale covariant generalizations of AdS and Schr\"odinger spacetimes. These embeddings rely on various aspects of dimensional reduction and flux compactification of eleven-dimensional supergravity.   Our top-down approach can be viewed as a partial holographic classification of the landscape of strongly coupled, UV complete quantum field theories with hyperscaling violation.

%%%
\Date{May  2012}
%%%%%%%%%%%%%%%%%%%%%%%%%%%%%%%%%%%%%%%%%%%%%%
%%%%%%%%%%%%%%%%%%%%%%%%%%%
% Main text begins here
%%%%%%%%%%%%%%%%%%%%%%%%%%%%%%%%%%%%%%%%%%%%%%
%%%%%%%%%%%%%%%%%%%%%%%%%%%
\baselineskip13pt

%\listtoc \writetoc

\newsec{Introduction}
Spacetime metrics that transform covariantly under dilatation have recently been re-interpreted as holographically dual to stress tensors of quantum field theories that violate hyperscaling \HuijseEF. Such a violation implies modified power law scaling of thermodynamic observables, relative to that of a conformal field theory. The purpose of this note is to present many supergravity solutions, old and new, with this property.

Much of the community's recent attention on this subject (e.g. \refs{\HuijseEF, \DongSE, \ShaghoulianAA, \HartnollWM,\SinghUN, \NarayanHK, \DeyTG, \DeyRS}) has gone into studying $d$-dimensional metrics of the form 
\eqn\ina{ds^2 = r^{-2\theta\over d-2}\left(-r^{2z}dt^2 + {dr^2\over r^2} + r^2 (dx^i)^2\right)}
where $i=1\ldots d-2$, which are scale covariant generalizations of metrics with Lifshitz symmetry \refs{\KachruYH,\KoroteevYP}: namely, \ina\ preserves translations and rotations, and has non-relativistic scale covariance with dynamical exponent $z$. The boundary of this spacetime lies at $r \rar\infty$. For $z=1$, this is Lorentz invariant, a scale covariant spacetime conformal to AdS. The parameter $\theta$, defined in Einstein frame, breaks scale invariance to covariance and maps holographically to what is known as the hyperscaling violation exponent. For a special value\foot{We warn the reader familiar with recent hyperscaling violation literature that we denote spacetime metrics as $d$-dimensional (or $D$-dimensional, depending on context), which interfaces more naturally with the supergravity literature; thus the ``hidden Fermi surface value'' of the hyperscaling violation exponent is $\theta=d-3$.} of $\theta$, \HuijseEF\ has argued that the bulk theory provides a holographic dual to a compressible state with a hidden Fermi surface. 

These metrics \ina, and finite temperature versions thereof, are solutions of the simple bottom-up Einstein-Maxwell-dilaton action \refs{\TaylorTG, \GubserQT,\CharmousisZZ, \PerlmutterQU}
\eqn\inb{S = \int d^dx \sqrt{-g} \left(R-{e^{\alpha\phi}\over 4}F_{\mu\nu}F^{\mu\nu} - \half(\p\phi)^2-V_0e^{\eta\phi}\right)}
Accompanying the running of the spacetime curvature scale is a rolling dilaton, and a gauge field which obeys a dilaton-dependent Gauss' law. From the top-down viewpoint, the metric \ina\ for $z=1$ is known to arise in the near-horizon geometries of D$p$-branes in type II supergravity; a handful of top-down embeddings of \ina\ with $z\neq 1$ and $\theta \neq 0$ have also recently been constructed in \refs{\SinghZS, \SinghUN, \NarayanHK, \DeyTG, \DeyRS}. 

There has been less, but some \KimNB, attention paid to a scale covariant form of the Schr\"odinger metric:
\eqn\inc{ds^2 = r^{-2\theta\over d-2}\left(-r^{2z}(dx^+)^2 + 2r^2dx^+dx^- +{dr^2\over r^2} + r^2 (dx^i)^2\right)}
where now, $i=1\ldots d-3$. This metric preserves translations, rotations and a Galilean boost symmetry, and when $\theta=0$ the full Schr\"odinger symmetry \refs{\SonYE, \BalasubramanianDM} is restored. The $\theta\neq 0$ metric has been embedded in supergravity in a limited number of cases -- TsT-transformed D$p$-branes \MazzucatoTR, and certain scaling limits of AdS solutions \SinghUN\ -- but has yet to be systematically understood as a solution to some gravitational action for many values of $z$, whether in bottom-up or top-down holography. The same can be said of the more general class of $D=d+n$-dimensional warped product metrics\foot{On the next page, we justify our use of the parameter $\theta$ in this higher-dimensional context.}
\eqn\incd{ds^2 = r^{-2\theta\over D-2}\left(r^{2z}q(y)(dx^+)^2 + 2r^2dx^+dx^- +{dr^2\over r^2} + r^2 (dx^i)^2 + ds^2_{{\cal{M}}_n}(y)\right)}
which involve an internal manifold ${\cal{M}}_n$ with dependence on coordinates $y$, and an arbitrary function $q(y)$; these leave the symmetries of \inc\ intact.  In an abuse of language, we will refer to these metrics as ``scale covariant Schr\"odinger'' metrics. 

In the following, we present supergravity embeddings in various dimensions of the metric \ina\ with $z=1$ and various values of $\theta$; the metric \inc\ with $(z=2, \theta = d-1)$; and infinite classes of metrics \incd, for many values of $(z,\theta)$. Some solutions are supersymmetric. 

An overarching theme is that dimensional reduction of theories admitting scale {\it invariant} vacua often leads to theories admitting scale {\it covariant} vacua. This statement is not original\foot{See e.g. \GouterauxQH\ for some recent related context.}, but we only wish to bring it to bear on the pursuit of classifying hyperscaling violation in field theory via AdS/CFT. Conversely -- and perhaps more usefully -- this gives us a way to think of the UV/IR completions of such solutions in terms of an uplift to a brane configuration at strong coupling. On the field theory side, this gives a concrete picture of the departure from the hyperscaling regime away from intermediate energy scales. Given the lack of pure field theory calculations at present on the allowed landscape of theories with hyperscaling violation, we encourage the view that gauge/gravity duality plays a powerful role in constraining this landscape \SachdevDQ; the work herein may be regarded as a step toward a full classification. 

The primary technical tool of this work is to map the spectrum of allowed pairs $(z,\theta)$ to harmonic spectra and cohomology of internal compactification manifolds. We will encounter tori, spheres, compact manifolds with non-trivial cohomology and combinations thereof.

Section 2 embeds the Lorentz invariant metrics \ina\ with $z=1$ in supergravity. In fact, these solutions are nothing but the near-horizon geometries of ``domain wall'' spacetimes of days of yore \LuRHB, ubiquitous in supergravity as members of a zoo of $p$-brane solutions. The latter are solutions to supergravities obtained from the maximal eleven-dimensional supergravity by dimensional reduction.  We review these results and translate into the language of hyperscaling violation, and give a partial catalog of the landscape of such solutions, both supersymmetric and non-supersymmetric.

In section 3, we turn to the metrics \inc\ and \incd. We write down a bottom-up Einstein-Maxwell-dilaton action which admits \inc\ as a solution: this action is nothing but \inb\ accompanied by a vector mass term. Our first top-down embedding is of a $(z=2, \theta=d-1)$ solution in the theory \inb\ with $V_0=0$ and $d\leq 10$, which arises in supergravity. More generally, we use a previous argument of \KrausPF\ to show how to generate infinite classes of such metrics by introducing null deformations of scale covariant supergravity solutions with a boost symmetry. 
To demonstrate our technique, we construct scale covariant Sch$_4$ solutions by null-deforming the type IIA extremal D2-brane geometry.

We conclude in section 4 with a discussion of holographic applications of these metrics, and some open questions. Appendix A includes some technical detail on our D2-brane solutions. \vs

\subsec{A prelude on dimensional reduction}
Before we begin, we wish to clarify the definition of $\theta$. Suppose we have some $D=d+n$-dimensional supergravity solution with Einstein frame metric conformal to a direct product,
\eqn\aba{ds^2_{D,E} = r^{-2X\over D-2}\left[ds^2_d + ds^2_{{\cal{M}}_n}\right]}
for some constant $X$, where ${\cal{M}}_n$ is some compact manifold and $ds^2_d$ is a scale invariant metric, e.g. one of the Lifshitz or Schr\"odinger metrics considered here. This metric will be accompanied in general by some running dilaton, so the theory in question can be written in non-Einstein frames. The parameter $\theta$ should be defined in Einstein frame, which enables proper contact with the dual field theory. 

We want to determine the relation between $X$ and $\theta$, where the latter dualizes to the hyperscaling violation exponent in the $(d-1)$-dimensional field theory. In fact, we now show that $X=\theta$, so that we can equally define $\theta$ in the decompactified $D$-dimensional solution. 

To see this, we make the traditional reduction ansatz
\eqn\abc{ds^2_{D,E} = e^{2\a\phi}ds^2_{d,E}+ e^{2\beta\phi}ds^2_{{\cal{M}}_n}}
where to reach the $d$-dimensional Einstein frame $ds^2_{d,E}$ requires
\eqn\abd{\beta=-{(d-2)\a\over n}}
Now, for the metric \aba\ one finds an Einstein frame metric
\eqn\abe{ds^2_{d,E} = r^{-2X\over d-2}ds^2_d}
which tells us that $X=\theta$ by definition: the factors one picks up from dimensional reduction and passage to the Einstein frame cancel to give $X=\theta$. 

Therefore, in what follows we use the parameter $\theta$ in both the $d$- and $D$-dimensional settings, and in application to product metrics warped by dependence on coordinates of the internal manifold ${\cal{M}}_n$ as in \incd.

\newsec{$z=1$ scale covariant metrics}

We first consider the simplest of the metrics with hyperscaling violation: the metric \ina\ with $z=1$, which is conformal to the Poincar\'e patch of AdS$_d$,
\eqn\dwa{\eqalign{ds^2 
 &= r^{-2\theta\over d-2} ds^2_{AdS_d}}}
This solves the action
\eqn\dwb{S = \int d^dx \sqrt{-g} \left(R - \half(\p\phi)^2-V_0e^{\eta\phi}\right)}
with 
\eqn\dwc{\eqalign{V_0 &= -[\theta-(d-1)][\theta-(d-2)]\cr
\eta^2 &= {2\theta\over (d-2)[\theta-(d-2)]}\cr   
\phi &= \eta\left(\theta-(d-2)\right)\log r}}
and $\theta$ finite, where reality and finiteness of $\eta$ -- alternatively, the null energy condition -- demand $\theta\leq 0$ or $\theta > d-2$. We have set the spacetime curvature radius to one. There are two more solutions of this theory of present interest: one is a finite temperature version of \dwa\ for which the definitions \dwc\ are identical, and which has a smooth $T\rar 0$ limit only when $\theta < 0$; and the other is an asymptotically flat version of \dwa. 
The latter geometries are referred to as ``domain walls'' in older literature because they describe $(d-2)$-branes, and one can view the geometry \dwa\ as the near-horizon limit of these domain walls \refs{\LuCS,\LuRHB}.

In the holographic condensed matter context, it was observed in \refs{\PerlmutterQU, \DongSE} that the action \dwb\ is the effective action for dimensionally reduced D$p$-branes in flat space, where $d=p+2$. In fact, the action \dwb\ can be obtained more generally in string theory, by consistent truncation of various supergravity theories in diverse dimensions.\foot{A subset of these solutions was recalled recently in \GouterauxCE.} This is simply a rejuvenation of old results. There are two ways of generating the action \dwa\ that we now recall: sphere reductions, and Scherk-Schwarz reductions. 

\subsec{Sphere reductions}
Our method will be to start with the maximal eleven-dimensional supergravity, and dimensionally reduce our way to \dwb. This was originally worked out in many papers, and there is a vast literature; we found \refs{\LuYN,\LuHB} to provide particularly helpful overviews.

First, we can toroidally reduce the maximal eleven-dimensional supergravity to obtain some supergravity in $D<11$ with a zoo of field strengths and scalars. The latter exponentially couple to the former, and both increase in number with the toroidal dimension via the Kaluza-Klein ansatz. Such a reduced supergravity will admit $p$-brane solutions - that is, $(p+1)$-dimensional extended geometries supported by flux -- which uplift to $D=11$ geometries of varying complexity \refs{\LuCS,\LuYN}. 
These $p$-branes are charged only under some subset of field strengths in the $D$-dimensional theory, analogous to the RR charges of D$p$-branes in type II string theory; as such we can truncate the theory to include just those field strengths, and whatever scalars to which they couple. This gives an action for the metric, $(11-D)$ scalar fields which descend from the eleven-dimensional metric, and $N$ field strengths of rank $n$ to which they couple:

\eqn\dwd{S = \int d^Dx \sqrt{-g} \left(R-\half(\p\vec{\phi})^2- {1\over 2n!}\sum_{\alpha=1}^{N}e^{\vec{c}_{\a}\cdot\vec{\phi}}(F_{n}^{\a})^2\right)}
where $\vec{c}_{\a}$ are constant $(11-D)$-vectors. 

The theory \dwd\ admits electric $(n-2)$- and magnetic $(D-n-2)$-brane solutions in which the $N$ field strengths can all carry distinct charges. The allowed pairs $(N,n)$ are determined by the dimension $D<11$ and the Kaluza-Klein procedure. The consistency of this truncation fixes the dot product between two $c_{\a}$; physically, we acquire this constraint on demanding that $p$-brane solutions to the full toroidally reduced supergravity are also solutions of \dwd.

Now, we can further truncate the action to include only a single $n$-form field strength, and a particular linear combination of scalars which we relabel $\vphi$:

\eqn\dwe{S = \int d^Dx \sqrt{-g} \left(R-\half (\p\vphi)^2-{1\over 2n!}e^{\a\vphi}(F_{n})^2\right)}
where the dilatonic coupling constant $\a$ is
\eqn\dwf{\a^2 = \Delta-{2(n-1)(D-n-1)\over D-2}}
for some constant $\Delta$. The $p$-brane solutions of this action are obtained from those of \dwd\ by taking all $N$ charges equal. For the action to admit {\it supersymmetric} $p$-brane solutions, we must have $\Delta=4/N$, where $N=1,2,3,4$ and $2^{-N}$ of the maximal supersymmetry is preserved. The effective actions for fundamental D$p$-brane solutions correspond to $\Delta=4$. There also exist non-supersymmetric solutions with other values of $\Delta$ which we will discuss momentarily. 

Following \CveticPN, we can now perform a consistent truncation on the $n$-sphere threaded by magnetic flux. 
This introduces a breathing mode scalar, but after further truncation one obtains the desired $d=(D-n)$-dimensional Lagrangian \dwb\ with only a single scalar, with 
\eqn\dwg{\eta^2 = \widehat{\Delta}  + 2{(d-1)\over d-2}}
where 
\eqn\dwh{\widehat{\Delta} = -{4(n-1)\Delta\over 2n\Delta-4(n-1)}} 
We have made contact with some older notation in which $\eta$ is parameterized not by $\theta$, but by $\widehat{\Delta} $, which is a quantity preserved under toroidal reduction. From \dwc, the relation between the two is
\eqn\dwi{\theta = D-n-2+{2\over \widehat{\Delta} +2}}

Similarly, we can perform a consistent truncation of \dwe\ for an {\it electric} flux configuration, in which case we end up with a $d=n$-dimensional action of the form \dwb; the expression for $\widehat{\Delta} $ is as in \dwh, but with $n\rar D-n$.

Any solution of one of these supergravity-embedded actions oxidizes to some M-theory configuration of intersecting branes, waves, and geometry, the rules for which are known explicitly \LuHB.\foot{We have been quick in our recollection of this chain of logic, but suffice it to say that an extremely rich set of actions and solutions has been unearthed of which we will not make use. The reader is referred to \refs{\DuffHP,\LuNT,\LuHB} and references therein for a robust sampling and classification of the many interesting solitonic solutions, both extremal and near-extremal, to toroidally reduced supergravities.} Historically, this has been a rather efficient way to find solutions which uplift to complicated M-theory configurations. Therefore, finding scale covariant solutions \dwa\ in supergravity boils down to asking what values of $\theta$ are produced by this series of reductions and truncations. This in turn amounts to classifying allowed triplets $(D,n,\Delta)$.

\subsec{Classification of solutions from sphere reduction}

First, let us contextualize two familiar supersymmetric examples. 

The cases of type II, 1/2-BPS D$p$-branes in $D=10$ reduced on the spheres arising in their near-horizon geometries give scale covariant metrics in $d=p+2$ dimensions, with 
\eqn\dwj{\theta={(p-3)^2\over p-5}}
These solutions are sourced by a single RR flux, so we take $\Delta=4$. For magnetic branes we take $n=8-p$ (reducing on $S^n$), and for electric branes, $n=p+2$. Plugging these values of $(D,n,\Delta)$ into \dwh\ and \dwi\ (or their electric counterparts) indeed gives \dwj. Taking $D<10$ instead just toroidally compactifies the D$p$-branes along their worldvolumes.

Another useful solution to keep in mind is that of the $D=6$ extremal, self-dual dyonic string, the near-horizon geometry of which is AdS$_3\times S^3$. This is nothing but the D1-D5 brane intersection 
reduced on $T^4$. So we will obtain a $\theta=0$ solution if we compactify on $S^3$. The solution is sourced by dyonic three-form flux, and the effective action of this dyonic string is just \dwe\ with $a=0$, the action for AdS$_3$ gravity: we have $D=6$, $\Delta=2$ (i.e. 1/4-BPS), and $n=3$, which does give $\theta=0$. \vs

Now we classify all values of $\theta$ consistent with the dimensional reduction to \dwb, starting with supersymmetric solutions. As we noted earlier, these must have $\Delta=4/N$, and preserve $2^{-N}$ of the maximal supersymmetry; for each $N$, one then looks for all pairs $(D,n)$ allowed by the reality of $\a$ (cf. \dwf). In fact, one can check that the only such pairs which give a finite $\theta\neq 0$ are the toroidal compactifications of D$p$-branes in flat space, described above. Certain pairs which give infinite $\theta$ are analogous to D5-branes which do not admit \dwa\ as a near-horizon geometry. Thus, in search of novel solutions we are forced to examine the spectrum of non-supersymmetric solutions, which need not have have $\Delta= 4/N$.

All values of $\Delta$ which arise for non-supersymmetric solutions obtained by sphere reduction are classified in \LuYN\ (see sections 4 and 5 therein). With respect to holographic applications to field theories which exhibit hyperscaling violation, we wish only to consider geometries with $d=D-n\geq 3$. All non-supersymmetric values of $\Delta$ for $d\geq 3$ are given\foot{This is exactly analogous to Table 2 of \LuYN, only we compile the non-supersymmetric rather than supersymmetric solutions.} in Table 1: 

\bigskip
\vbox{
$$\vbox{\offinterlineskip
\hrule height 1.1pt
\halign{&\vrule width 1.1pt#
&\strut\quad#\hfil\quad&
\vrule width 1.1pt#
&\strut\quad#\hfil\quad&
\vrule width 1.1pt#
&\strut\quad#\hfil\quad&
\vrule width 1.1pt#\cr
%&\strut\quad#\hfil\quad&
%\vrule width 1.1pt#\cr
height3pt
&\omit&
&\omit&
&\omit&
%&\omit&
\cr
&\hfil $ $ &
&\hfil $n=3$&
&\hfil $n=2 $&
%&\hfil $n_H$ &
\cr
height3pt
&\omit&
&\omit&
&\omit&
%&\omit&
\cr
\noalign{\hrule height 1.1pt}
height3pt
&\omit&
&\omit&
&\omit&
%&\omit&
\cr
&\hfil $D=9$ &
&\hfil $3$&
&\hfil $3 $&
%&\hfil $..$&
\cr
height3pt
&\omit&
&\omit&
&\omit&
%&\omit&
\cr
\noalign{\hrule}
height3pt
&\omit&
&\omit&
&\omit&
%&\omit&
\cr
&\hfil $D=8$ &
&\hfil ${8\over 3}$&
&\hfil ${8\over 3},{12\over 7} $&
%&\hfil $..$&
\cr
height3pt
&\omit&
&\omit&
&\omit&
%&\omit&
\cr
\noalign{\hrule}
height3pt
&\omit&
&\omit&
&\omit&
%&\omit&
\cr
&\hfil $D=7$ &
&\hfil ${5\over 2}$&
&\hfil ${5\over 2},{5\over 3},{8\over 5}$&
%&\hfil $..$&
\cr
height3pt
&\omit&
&\omit&
&\omit&
%&\omit&
\cr
\noalign{\hrule}
height3pt
&\omit&
&\omit&
&\omit&
%&\omit&
\cr
&\hfil $D=6$ &
&\hfil ${12\over 5}$&
&\hfil ${3\over 2},{12\over 5},{8\over 5},{20\over 13}$&
%&\hfil $..$&
\cr
\noalign{\hrule}
height3pt
&\omit&
&\omit&
&\omit&
%&\omit&
\cr
&\hfil $D=5$ &
&\hfil $-$&
&\hfil ${24\over 17},{7\over 3},{3\over 2},{15\over 11},{7\over 5}, {4\over 3}$&
%&\hfil $..$&
\cr
%\noalign{\hrule}
%height3pt
&\omit&
&\omit&
&\omit&
%&\omit&
\cr
}\hrule height 1.1pt
}
$$
}
\centerline{\sl Table 1: Non-supersymmetric values of $\Delta$, for $D-n\geq 3$. }\bigskip
%%%%%%%%%%%%
Once an entry appears for given $D$ at fixed $n$, its appearance is also implied for all lower $D$. One can easily translate this table into values of $\theta$ using \dwh\ and \dwi. For the sake of application to physical systems of interest, we present all finite values of $\theta\neq 0$ one can obtain in a non-supersymmetric solution in $d=4,5$:

\eqn\dwl{\eqalign{d=4:&\quad \theta=-3, -2, -1, -{1\over 3}, 6, 7, 8, 9\cr
d=5:& \quad \theta = -2,-1,7,8,9\cr}}
These parameterize hyperscaling violation in $(d-1)$-dimensional quantum field theories. 

We note in passing that, of course, these supergravity solutions obey the null energy condition, which requires $\theta \leq 0$ or $\theta > d-2$  \DongSE. We emphasize that each of these solutions has a known uplift to an eleven-dimensional geometry, and we refer the reader to previous references for details. The ability to identify the worldvolume field theories of the uplifted brane configurations should be a useful tool for studying hyperscaling violation in UV complete field theories with supergravity duals.

\subsec{Scherk-Schwarz reductions}
One can also obtain the action \dwb\ in the following way. Starting from some higher-dimensional supergravity, one makes a Kaluza-Klein ansatz in which some field strength is a linear combination of harmonic forms on some Ricci-flat compact manifold ${\cal{M}}$ with non-trivial cohomology. The reduction is consistent as long as the potential appears in the action only via its field strength. This is referred to as a ``generalized dimensional reduction'' in \LavrinenkoMP, or Scherk-Schwarz reduction \ScherkZR.

When ${\cal{M}}$ is a torus, one can engineer various complicated intersecting brane configurations which are supported by $N$ distinct charges. To extend this to more general classes of ${\cal{M}}$, notably Calabi-Yau manifolds, one must simplify the ansatz by taking all charges equal; as we know from our previous discussion on sphere reductions, this has the effect of generating an effective action with a single-scalar exponential potential, namely \dwb\ with
\eqn\dwq{\eta^2 = \widehat{\Delta}  + 2{(d-1)\over d-2}}
and
\eqn\dwrs{\widehat{\Delta} = {4\over N}~, ~~ 2\leq N \leq 8}
Again translating to the hyperscaling violation exponent $\theta$, we have
\eqn\dwt{\theta=d-2+{2N\over 2N+4}}
The associated solutions \dwa\ preserve half of the $d$-dimensional supersymmetry, itself a function of the compactification manifold ${\cal{M}}$ \LavrinenkoMP. For instance, solutions \dwa\ in reductions of M-theory on ${\cal{M}}=CY_3$ preserve four real supercharges. 

The pairs $(d,N)$ which can be realized in supergravity are a function of cohomology and the spectra of field strengths. This is a vast landscape which we will not fully classify here; the reader is referred to \LavrinenkoMP\ for a more methodical exposition and several examples. 

For practical purposes, let us quote from \LavrinenkoMP\ three values of $\theta$ one can obtain in $d=4$ by this method:
\eqn\dwu{d=4:\quad \theta= {13\over 5}, {8\over 3}, {25\over 9}}
which are $N=3,4,7$ solutions, respectively. The $N=4$ solution, for example, is constructed as follows. One begins with type IIA supergravity in $D=10$ and turns on components of $F_3$ proportional to a particular harmonic form of $CY_3$. There is a solution describing four intersecting 5-branes charged under $F_3$, each of which is wrapped on a harmonic 3-cycle in $CY_3$. Upon taking the charges equal and reducing to $d=4$, their mutual intersection gives rise to a ``domain wall'' geometry, the near-horizon of which is the scale covariant metric \dwa\ with $\theta={8\over 3}$.

\newsec{Scale covariant Schr\"odinger metrics}
The metrics we consider here are \inc,
\eqn\incg{ds^2 = r^{-2\theta\over d-2}\left(-r^{2z}(dx^+)^2 + 2r^2dx^+dx^- +{dr^2\over r^2} + r^2 (dx^i)^2\right)}
and \incd,
\eqn\scha{ds^2 = r^{-2\theta\over D-2}\left(r^{2z}q(y)(dx^+)^2 + 2r^2dx^+dx^- +{dr^2\over r^2} + r^2 (dx^i)^2 + ds^2_{{\cal{M}}}(y)\right)}
These preserve rotations, translations and Galilean boosts; at $\theta=0$, these are the Schr\"odinger metrics first introduced in \refs{\SonYE, \BalasubramanianDM}. Despite the fact that we cannot reduce over ${\cal{M}}$ in \scha, we continue to define $\theta$ in the above manner as discussed in the introduction.\foot{One may think of the $x^-$ direction as noncompact, in the spirit of \refs{\GuicaSW,\SongSR}, in which case the dual QFT is nonlocal in the $x^-$ direction and looks like a null dipole theory; our solutions here then describe the breaking of scale invariance to covariance in the null dipole theory. One may also compactify this coordinate in keeping with a particle number interpretation, with the caveat that light wound strings threaten the validity of the solution \MaldacenaWH.}

For $\theta=0$, this class of metrics is straightforwardly embedded into string and M-theory (e.g. \refs{\MaldacenaWH, \HartnollRS, \DonosEN, \BobevMW, \DonosXC, \JeongAA, \KrausPF}), and we will soon extend those methods to the $\theta \neq 0$ case. Before doing so, we present a bottom-up action of an Einstein-Maxwell-dilaton theory which admits the simpler solution \incg, and in fact appears in supergravity for certain values of parameters. 
\subsec{Bottom-up}
Recall that the breaking of AdS scale invariance to covariance simply requires adding a scalar field with an exponential potential to the AdS action: the linear dilaton background generates power law behavior consistent with scale covariance. One can think of this as AdS with an $r$-dependent radius of curvature, valid only at intermediate radii. 

In analogy with this, we can generate scale {\it covariant} Schr\"odinger solutions ($\theta\neq 0$) by adding a scalar with exponential potential to the theory of a massive vector coupled to gravity, which is known to give scale {\it invariant} Schr\"odinger solutions ($\theta= 0$). That is, the massive vector action
\eqn\mva{S = \int d^dx \sqrt{-g} \left(R-{1\over 4}F_{\mu\nu}F^{\mu\nu} - {m^2\over 2}A_{\mu}A^{\mu} - 2\Lambda \right)}
admits a Schr\"odinger solution \refs{\SonYE, \BalasubramanianDM} for parameters
\eqn\mvb{m^2=z(z+d-3)~, ~~ -2\Lambda={(d-1)(d-2)}}
where we set the spacetime radius to one. In a supergravity context, it was shown in \KrausPF\ that these vectors are simply KK modes of an AdS$\times {\cal{M}}$ solution. So, to admit a $\theta\neq 0$ solution, we add a real scalar with an exponential potential. We also should add a running gauge coupling. Thus, we consider the $d$-dimensional action
\eqn\sca{S = \int d^dx \sqrt{-g} \left(R-{e^{\alpha\phi}\over 4}F_{\mu\nu}F^{\mu\nu} - {m^2\over 2}A_{\mu}A^{\mu} - \half(\p\phi)^2-V_0e^{\eta\phi}\right)}

One can view this action as just the addition of a vector mass term to the Lagrangian \inb, or the addition of a rolling potential and gauge coupling to the massive vector theory \mva. The latter is effectively the action for a KK vector of the relativistic scale covariant metrics of the previous section.\foot{This is in the general class of theories considered in \aprile, where the authors construct holographic superconductors; however, they did not consider the exponential potentials that give rise to hyperscaling violating solutions, instead tuning the scalar couplings to allow asymptotically AdS black holes.} 

Our ansatz is
\eqn\hyh{\eqalign{A &= ar^z dx^+\cr
\phi&=C\log r\cr
ds^2&= r^{-2\theta/(d-2)}\left(-\sigma^2r^{2z}(dx^+)^2+ 2r^2dx^+dx^- + {dr^2\over r^2} + r^2(dx^i)^2\right)\cr}}
where $i=1\ldots d-3$. We have shifted away a possible constant in $\phi$ by shifting $r$. In addition to the parameter $\sigma$ which we could scale away, the ansatz contains four parameters $(a,z,C,\theta)$, as does the Lagrangian which is a function of $(\alpha,\eta,m,V_0)$. 

The vector is null on account of $g^{++}=0$, so $F^2=A^2=0$ on-shell. When $\sigma=0$, this is a relativistic ansatz, and we have $a=C=0$. When $\eta=\alpha=\phi=0$, we recover the massive vector theory, and a Schr\"odinger solution exists with relations \mvb\ and $a^2 =  2\sigma^2(z-1)/ z$. The fact that $a\propto \sigma$ implies that $\sigma$ is dual to a constant source for a null vector operator in the action of the putative dual CFT \GuicaSW. 

The field equations are
\eqn\hyj{\eqalign{\square \phi &= V_0\eta e^{\eta\phi}\cr
\p_{\mu}(\sqrt{g}e^{\alpha\phi}F^{\mu\nu}) &= m^2\sqrt{g}A^{\nu}\cr
G_{\mu\nu} &= {e^{\alpha\phi}\over 2}F_{\mu\nu}^2 + {m^2\over 2}e^{\chi\phi}A_{\mu}A_{\nu}+\half\p_{\mu}\phi\p_{\nu}\phi\cr&-{1\over 4}\left[(\p\phi)^2+2V_0e^{\eta\phi}\right]g_{\mu\nu}\cr}}
where we have dropped terms $A^2=F^2=0$, as these vanish on our ansatz. We find that the ansatz \hyh\ is a solution with the following simple relations:
\eqn\scc{\eqalign{V_0 &= -[\theta-(d-1)][\theta-(d-2)]\cr
\eta^2 &= {2\theta\over (d-2)[\theta-(d-2)]}~, \quad \alpha   = -\eta \cr
C &= \eta(\theta-(d-2))\cr
m^2 &= z(z+d-3-\theta)\cr
a^2 &= {2\sigma^2 (z-1)\over z}\cr}}
The solutions leave three free parameters: $(\sigma,z,\theta)$. These manifestly reduce to the scale invariant Schr\"odinger solution in the $\theta=0$ limit, and to the $z=1$ scale covariant solution.
 
Notice that the parameters $(V_0,\eta,C)$ have the same definitions as in \dwc, as expected by analogy with the relation between Schr\"odinger and AdS solutions in the massive vector theory \mva.  So again, we have bounds
\eqn\scd{\theta \leq 0 \quad {\rm or} \quad \theta > d-2}
but now reality of $m$ and $a$ then further constrain values of $z$ in each of these regions of $\theta$:
\eqn\sce{\eqalign{\theta \leq 0:&\quad z \geq 1 ~ {\rm or} ~ z \leq -(d-3-\theta)\cr
\theta > d-2:&\quad z \geq 1 ~ {\rm or} ~ z <0\cr}}

The fact that $a\propto \sigma$ once again tells us that we have turned on a null vector operator in the field theory dual with constant source $\sigma$.

Let us point out that there are nontrivial $m=0$ solutions here: at $\theta=z+d-3$, we have $m=0$, and these solutions are consistent with \scd\ for both $z\leq 3-d$ and $z > 1$. This is an unexpected result: the same Einstein-Maxwell-dilaton system which admits the much-studied scale covariant Lifshitz metrics also admits a certain class of scale covariant Schr\"odinger metrics.\foot{We also note that while we are aware of string theory embeddings \CveticPU\ of the Lagrangian \sca\  which have $m=0$ and $\eta=-\a=\sqrt{2\over d-2}$, this value is inconsistent with finite $\theta$. Perhaps there are other embeddings of \sca\ unknown to us.}
 
\subsec{Top-down, I.}

When $V_0=m=0$ in the action \sca, we recover our supergravity action \dwe\ (where $D_{there}=d_{here}$). This happens when 
\eqn\scea{\theta=d-1~, ~~ z=2~,~~\a^2 = {2(d-1)\over d-2}}

The values of $\a$ which appear in supergravity were given in \dwf; one easily checks that $\a^2={2(d-1)\over d-2}$ when $\Delta=4$ . Now, actions \dwe\ with two-form field strengths and $\Delta=4$ do indeed exist for all $d\leq 10$: for any $p$-form field strength, the value $\Delta=4$ obtains when the field strength in the action descends directly from an eleven-dimensional field rather than being a linear combination of different field strengths. The first two-form field strength appears in $d=10$ as the Kaluza-Klein gauge field of the eleven-dimensional metric. Therefore, we have shown that the scale covariant Schr\"odinger solution  with $\theta=d-1, z=2$ exists in maximal supergravities of any dimension $d=3,4,\ldots 10$.  \vs

We now turn to another way to generate infinite classes of supergravity solutions with metric \scha.

\subsec{Top-down, II.}
In \KrausPF, the following argument was established. Consider a solution of some theory of gravity -- any supergravity, say -- which preserves a boost symmetry. Let us write the $D$-dimensional line element as
\eqn\pfa{ds^2 = g_{\mu\nu} dx^{\mu}dx^{\nu} = g_{+-}dx^+dx^- + g_{ij}dx^{i}dx^{j}}
where $i=1\ldots D-2$. The boost acts on the lightcone coordinates $x^{\pm}$ as
\eqn\pfb{x^+ \rar {x^{+}}'={x^+ \over \kappa}\, ,\quad x^- \rar {x^-}'=\kappa  x^-}
so we take $g_{+-}$ and $g_{ij}dx^{i}dx^{j}$ to be invariant under this boost. Let us also take the full solution to be invariant under shifts in $x^{\pm}$. 

Now, we seed a perturbation around this solution by turning on a linearized metric fluctuation. In particular, we take this fluctuation to lie along the $(++)$ direction:
\eqn\pfc{\delta g_{\mu\nu}^{(1)} = g_{++}^{(1)}\delta_{\mu +}\delta_{\nu +}}
Under the boost \pfb, the metric component $g_{++}$ has weight two, and it preserves shifts along $x^{\pm}$. One may think of this field as a linearized Kaluza-Klein mode.

The key point is that at $n^{th}$ order in perturbation theory, we are constrained to add field components which preserve the shift symmetry and have boost weight $2n$, because the $n=1$ metric perturbation acts as a source. In a theory with fields of spin less than or equal to two -- perhaps a theory of scalars, fluxes and gravity, as in tree-level supergravity -- there is no field which can give this behavior, even at $n=2$ order. Therefore, the linearized solution is in fact a nonlinear solution of the theory: that is, the metric
\eqn\pfcd{\widetilde{g}_{\mu\nu}  = g_{\mu\nu} +  g_{++}^{(1)}\delta_{\mu +}\delta_{\nu +}}
solves the full equations of motion for some $g_{++}^{(1)}$ consistent with remaining symmetries. The curvature scale of the spacetime will be unchanged, and the $x^+$ direction remains null to all orders. 

One can repeat this logic, instead starting with a linearized weight one perturbation of a field other than the metric: for instance, if we turn on a component of flux which has a leg along the $dx^+$ direction, the perturbation theory truncates at second order after having turned on the metric perturbation \pfc\ (now at second order). This second-order solution will exist according to the field equations of the theory.\foot{In the classical, large N limit, the existence of such solutions is independent of subtleties due to higher spin bulk fields dual to multi-trace operators. See \KrausPF\ for details.}

In the case that the zeroth order solution possesses a definite weight under scale transformation, the perturbation theory must be consistent with this too. 
Thinking of the bulk metric as dual to a quantum field theory, and writing the boundary coordinates as $(x^\pm, \xb)$, the generators corresponding to dilatations and Lorentz boosts in $x^\pm$
act as
\eqn\pfcb{\eqalign{{\cal D}:\quad &x^+ \rar  {x^{+}}'={x^+ \over \lambda}\, ,\quad x^- \rar {x^-}'= {x^- \over \lambda}\, ,\quad \xb \rar \xb'= {\xb \over \lambda}  \cr
{\cal L}:\quad  & x^+ \rar  {x^{+}}'={x^+ \over \kappa}\, ,\quad x^- \rar {x^-}'=\kappa  x^-\, ,\quad \xb \rar \xb'= {\xb }\cr}}
Suppose our zeroth order line element transforms covariantly under scaling:
\eqn\pfd{ds^2 \rar (ds^2) ' = \lambda^{-2\theta/(D-2)}ds^2}
with hyperscaling violation exponent $\theta$. The null deformed solution is no longer Lorentz invariant, so it is consistent to allow an anisotropic scaling between the $x^{\pm}$ directions that preserves the behavior \pfd: the metric now transforms covariantly under non-relativistic dilatations, generated by \GuicaSW\
\eqn\pff{{\cal{D}}_{z} = {\cal{D}} + (z-1){\cal{L}}}
which acts as
\eqn\pfg{\eqalign{{\cal D}_{z}:\quad &x^+ \rar {x^{+}}'= {x^+ \over \lambda^z}\, ,\quad x^- \rar {x^{-}}'={x^- \over \lambda^{2-z}}\, ,\quad \xb \rar \xb'= {\xb \over \lambda}}}
Upon further assuming translation and rotation invariance of the solution, the final metric is a scale covariant Schr\"odinger solution, still with the same value of $\theta$. 

To summarize, the construction of gravity solutions with Schr\"odinger symmetry with or without hyperscaling violation automatically follows from perturbation theory around a boost-symmetric solution along a lightcone direction. In particular, Schr\"odinger solutions are null deformations of AdS, and scale-covarant Schr\"odinger solutions are null deformations of scale covariant spacetimes \dwa. Knowledge of the linearized KK spectrum is essentially enough to determine the space of such solutions. Since the D$p$-brane spacetimes (in Einstein frame) are not direct products, one can think of the masses of KK fields as being $r$-dependent. This explains the success of the bottom-up Lagrangian of the previous subsection. 

One easy way, therefore, to generate infinite classes of Schr\"odinger solutions with hyperscaling violation is to null deform nonconformal brane spacetimes of string theory. Lest our argument appear too indirect, we now show this explicitly for D2-branes in flat space.

\subsec{Top down, IIA: A D2-brane example}

While we construct our scale covariant Sch$_4$ solutions directly in type IIA using D2-branes, the following solutions can in fact be viewed as scale {\it invariant} Sch$_4$ deformations of the conformal M2 brane geometry \DonosXC, smeared along the M-theory circle transverse to the branes and then dimensionally reduced to type IIA. We also note that the following construction of infinite families of scale covariant Sch$_4$ solutions using D2-branes in flat $\BR^7$ can be generalized upon replacing $\BR^7$ by a cone over any Einstein space with positive Ricci curvature.\foot{The method and notation are modeled directly on those of \refs{\DonosXC,  \KrausPF}.} 

The Einstein frame field equations and Bianchi identities for the nonzero bosonic fields $(g_{\mu\nu},F_4,\phi)$ in our type IIA supergravity ansatz are given in Appendix A. Our ansatz is
\eqn\dtb{\eqalign{e^{\phi} &= H^{1/4}\cr
F_4 &= dH^{-1}\wedge dx^+\wedge dx^-\wedge dx + dx^+\wedge W\cr
ds^2 
&= H^{-5/8}(h(dx^+)^2+2dx^+dx^-+dx^2)+H^{3/8}ds^2_{\BR^7}\cr}}
where $W$ is a real three-form and $(h,H)$ are real functions, all defined on $\BR^7$. One may think of this as a D2-brane with a worldvolume wave supported by null four-form flux. When $h=W=0$, this is the Einstein frame solution for D2-branes in flat space where $H$ is harmonic on $\BR^7$,
\eqn\dtc{d\star_{\BR^7} dH=0}
For $(h,W)$ nonzero, this is a solution if
\eqn\dtn{\eqalign{\nabla^2_{\BR^7} h &= -|W|^2_{\BR^7}\cr
dW&=0\cr
d\star_{\BR^7}W&=0\cr
d\star_{\BR^7}dH&=0}}
where $|W|^2_{\BR^7} = {1\over 3!}W_{abc}W^{abc}$. 

Now we wish to zoom into the near-horizon region and construct the Schr\"odinger solution with $\theta \neq 0$. Write $\BR^7$ as a cone over $S^6$:
\eqn\dto{ds^2_{\BR^7} = dr^2 + r^2d\Omega^2_6}
To zoom into the near-horizon and satisfy the harmonic condition on $H$, take $H=r^{-5}$. The Einstein metric reads
\eqn\dtp{ds^2 = r^{1/8}\left(r^{3}(2dx^+dx^- + h(dx^+)^2 + dx^2)+ {dr^2\over r^2} + d\Omega^2_6\right)}
To obtain a metric with non-relativistic scale covariance, take 
\eqn\dtpa{h=r^{3(z-1)}q}
where $q$ is a function on $S^6$. For most of our solutions, $q$ will not be constant. This metric has $\theta=-1/3$, as it must in accord with our perturbation theory arguments and \dwj.

We now must specify $W$. To ensure $dW=0$, take 
\eqn\dtpb{W=d(r^x\tau)}
for some constant $x$, where $\tau$ is a two-form on $S^6$. By the first equation in \dtn, the left-hand side scales like $r^{3z-5}$; a dilatation fixes $x={3z\over 2}+\half$. This is also the value one obtains by requiring that the flux $F_4$ should scale with some fixed power of $r$, and examining the scaling weight of the D2 background term. 

Given these expressions \dtpa\ and \dtpb, we plug into the equations \dtn. The result is a linear eigenvalue equation for co-closed two-forms on $S^6$, and an inhomogeneous Laplace equation for scalars on $S^6$:
\eqn\dtu{\eqalign{&\Delta_{2}\tau = \left({3z\over 2}+\half\right)\left({3z\over 2}+{3\over 2}\right) \tau~, \quad {\rm where}~ d\star_{S^6}\tau=0\cr
&\nabla^2q + (3z-3)(3z+2)q = -\left[\left({3z\over 2}+\half\right)^2|\tau|^2_{S^6} + |d\tau|^2_{S^6}\right]\cr}}
where we have defined the Laplace operator acting on transverse two-forms $\tau$ as $\Delta_2 =d^{\dagger}\cdot d + d\cdot d^{\dagger}$, and each of these Laplace operators is understood to be acting on forms defined on $S^6$. Finding a solution boils down to harmonic analysis on $S^6$.

Using the shorthand $Y_p^{\ell}$ to denote the $p$-form transverse harmonics on $S^6$ of degree $p+l$ in the basic scalar harmonic (see Appendix A for details), the harmonic equations for $p=0,2$ are
\eqn\dug{\eqalign{\nabla^2Y_0^{\ell} &= -\ell(\ell+5)Y_0^\ell~, \quad \ell=1,2,\dots\cr
\Delta_2Y_2^n &= (n+2)(n+3)Y_2^n~, \quad n=1,2,\dots\cr}} with eigenvalue degeneracies listed in appendix A.

Returning to the system \dtu, there are two types of solutions:\vs

\bul When $\tau=0$, this is just a linearized spin-2 perturbation of the sort described in the previous subsection: the field equation for $q$ is precisely that of a spin-2 KK mode with $(mL)^2 = (3z-3)(3z+2)$. \dug\ tells us that $q$ is simply a scalar harmonic $Y_0^\ell$, where $\ell$ parameterizes an infinite family of solutions. There are two branches of solutions -- one for positive $z$ and one for negative\foot{The $z=-{2\over 3}$ solution was discovered in \SinghUN\ by a different method.} $z$ -- and we list the positive branch only:
\eqn\dtw{\eqalign{z&=1+{\ell\over 3} = {4\over 3},{5\over 3},\ldots\cr}}
These solutions constitute an extension of those first constructed in \HartnollRS\ to the scale covariant case. Following \refs{\HartnollRS,\BobevMW,\DonosXC,\BobevQX} we expect them to preserve one quarter of the pre-existing supersymmetry, which in this case gives a 1/8-BPS solution; it would obviously be worthwhile to confirm this directly.  \vs

\bul When $\tau\neq 0$, the KK vector perturbation turns on spin-2 modes as well. First, we consider the eigenvalue equation for $\tau$ which determines $z$ as (taking the positive branch)
\eqn\dty{z = 1+{2n\over 3} = {5\over 3},{7\over 3},\ldots}
Now, note that for all of these values of $z$, there exists a homogeneous solution to the scalar Laplace equation in \dtu; that is, given \dty, we can write the equations \dtu\ in terms of the eigenvalue of the two-form harmonic as
\eqn\dtua{\eqalign{&\Delta_{S^6}\tau = \left(n+2\right)\left(n+3\right) \tau~, \quad {\rm where}\quad d\star_{S^6}\tau=0\cr
&\nabla^2_{S^6}q + 2n(2n+5)q = -\left[\left(n+2\right)^2|\tau|^2_{S^6} + |d\tau|^2_{S^6}\right]\cr}}
Thus, in order for there to be a solution to the second equation for a given $n$, we must show that 
with $\tau$ a transverse two-form $\tau\sim Y_2$, the object
\eqn\duj{\Lambda(n) \equiv (n+2)^2|\tau|_{S^6}^2 + |d\tau|_{S^6}^2~, \quad n=1,2,\ldots}
does not source the scalar harmonic $Y_0^{\ell=2n}$.

In appendix A, we show explicitly that this is true for the lowest two-form, $n=1$. This corresponds to a solution with $z=5/3$. To verify that the $Y_0^{\ell=2}$ harmonic is not sourced by the quadratic matter terms, we need to manipulate the $Y_2^{n=1}$ harmonic in $S^6$. The result is that
\eqn\dtza{\Lambda(n=1) = \left[9|\tau|^2_{S^6} + |d\tau|^2_{S^6}\right] = 9}
up to overall normalization of $\tau$. This involves nontrivial cancellations. Therefore, there is a solution to \dtua\ with $q$ a negative constant:
\eqn\dupa{q = -{9\over 14}}
This is the correct sign needed for us to interpret $x^+$ as the time coordinate of a dual field theory. 

To this, we can add an arbitrary linear combination of $Y_0^{\ell=2}$ harmonics; additionally, we could have taken $\tau$ to be a linear combination of $Y_2^{n=1}$ harmonics, and $q$ would remain a negative constant. In all, we have (at most) a 62-dimensional family of such solutions, and a 35-dimensional family of solutions with $q$ constant. The $n=1$ solution with $q$ constant is particularly amenable to development of an AdS/CFT dictionary, as one can reduce on the $S^6$ and view the geometry in a four-dimensional light. 

Further discussion of technical details is relegated to Appendix A. We only note here that for an ansatz built on the $n^{th}$ harmonic two-form $\tau$ (cf. \dtua), it is natural to expect that the quadratic source $\Lambda(n)$ only turns on the $\ell=2n-2$ scalar harmonic, based on the above and on previous work with scale invariant Schr\"odinger solutions. (See, e.g., section 3 of \BobevMW.) A proof of this is probably straightforward, perhaps using techniques in \DonosXC. 
\vs
To summarize, we have good evidence that a solution to \dtua\ exists for all integer $n\geq 1$, and hence that the scale covariant Sch$_4(z=1+{2n\over 3})$ solution of type IIA supergravity exists for all such $n$. We have shown this explicitly for $n=1$ only. These are in addition to the $\tau=0$ solutions, for which an infinity of scale covariant Sch$_4(z=1+{\ell\over 3})$ solutions exists for all integer $\ell\geq 1$. In fact, by viewing these solutions as dimensionally reduced\foot{It is straightforward to verify this claim, for instance by restricting the fields of the M-theory ansatz to be independent of the eleventh dimension.} Sch$_4$ solutions of M-theory, first constructed in \DonosXC\ by an M2-brane null deformation, we know that the full spectrum of solutions must exist and their supersymmetries are largely understood \refs{\DonosXC,\DonosZF}. We emphasize that this is but one example of the algorithm recalled in the previous subsection for generating solutions with Galilean boost symmetry.

\subsec{Comments}
These solutions are satisfying because they have known resolutions of their IR and UV singularities. In general, scale covariant supergravity solutions are valid only over a limited range of energy scales \ItzhakiDD. For relativistic D2 branes, we know that the dual field theory flows to fixed points in the IR (a conformal Chern-Simons-matter theory) and the UV (free super Yang-Mills). Similarly, the field theory duals to the solutions we have constructed here flow to scale invariant, non-relativistic deformations of the M2 worldvolume theory in the IR, dual to the solutions in \DonosXC; they should become essentially trivial in the UV where the Yang-Mills theory becomes free.

Generally speaking, one can dimensionally reduce various supergravity Schr\"odinger solutions in the literature to obtain scale covariant metrics with Galilean boost symmetry.

\vs

\newsec{Ruminations and applications}
To conclude, we have reviewed and enlarged the family of known supergravity solutions which display scale covariance.  These map to gauge theories living on branes and their intersections which violate hyperscaling at strong 't Hooft coupling. Given the paucity of results on hyperscaling violation in field theory, this is one of the exciting situations in holography in which we are not merely extrapolating known field theory results to strong coupling. As with other string and M-theory embeddings of gravity duals to condensed matter phenomena -- for instance, in analogy to the progression of work on holographic superconductors (e.g. \refs{\HartnollVX, \DenefTP, \GubserQM, \GauntlettDN, \BobevRV}) -- we hope that the present work will be viewed as one basic ingredient in the holographic study of hyperscaling violation in fully {\it consistent} quantum field theories. This should enable holographic study of hyperscaling to move beyond an effective approach and, with microscopic understanding, address the field theory dynamics in the hyperscaling regime and near the crossover scale. 

The major class of embeddings missing from this paper are the scale covariant Lifshitz solutions \ina\ with $z\neq 1$, so we have nothing to say about geometries which may (or may not \HartnollWM) hide Fermi surfaces \HuijseEF. It would be very useful to have supergravity embeddings of the action \inb. It may be more profitable to take a less direct approach, more along the lines of our section 3, and use the full power of ten- or eleven-dimensional geometry to find these Lifshitz solutions; this was the tack of \refs{\DonosTU,\SinghUN, \NarayanHK, \DeyTG, \DeyRS}. It seems feasible that a generalization of the arguments in \CassaniSV\ can be made to identify scale covariant Lifshitz solutions in generic theories including multiple scalars. 

We also did not discuss the stability of our solutions here; it is good to keep in mind that as argued in \DongSE, our $z=1$ solutions with $\theta$ greater than the number of field theory spatial dimensions may be unstable, at least thermally if not otherwise. This has precedent in studies of the decoupling limit of D$p$-branes with $p>5$ and the question of existence of an attendant holographic correspondence \BoonstraMP. 

Still on the subject of stability, many of the solutions presented here are not supersymmetric, which is desirable from the boundary point of view but comes with the usual gravitational stability caveat. For those that are supersymmetric, it would be nice to confirm and study their supersymmetry explicitly in ten-dimensional language. 

Finally, as for a Fermi surface interpretation of scale covariant Schr\"odinger solutions, the reference \KimNB\ studied the entanglement entropy in the background \inc\ and found area law violations \OgawaBZ. We did not realize this parameter range with any of our supergravity embeddings; furthermore, for the more general null-deformed ten- and eleven-dimensional supergravity solutions \incd\ that we constructed in section 3, the analysis of \KimNB\ does not apply. Given that more UV complete holographic duals are now available, one is well equipped to study the issue of hidden Fermi surfaces directly in supergravity. 

\bigskip
\noindent {\bf Acknowledgments:} \medskip \noindent 

We thank Martin Ammon, Shamit Kachru, Per Kraus and K. Narayan for helpful discussions, as well as the Erwin Schr\"odinger Institute for hospitality during the later stages of this work. The author is supported in part by a Dissertation Year Fellowship from the UCLA Graduate Division.

\appendix{A}{Details of scale covariant Sch$_4$ solutions from D2 branes}
\subsec{Type IIA field equations}
The type IIA supergravity field equations in Einstein frame for $(g_{\mu\nu},F_4,\phi)\neq 0$ and all other fields zero are as follows:
\eqn\dta{\eqalign{d\left(e^{\phi/2}\star F_4\right)&=0~, ~~ F_4\wedge F_4=0~, ~~d F_4=0 \cr
d\star d\phi &= {1\over 4}e^{\phi/2}F_4\wedge \star F_4\cr
R_{\mu\nu} &= \half \p_{\mu}\phi\p_{\nu}\phi+{1\over 12}e^{\phi/2}F_{\mu\nu}^2-{1\over 128}e^{\phi/2}F^2g_{\mu\nu}}}
\subsec{Scalar and two-form harmonics on $S^6$}
Our treatment here takes after the nice exposition in \refs{\BobevMW, \BobevQX}, and  \ElizaldeNB\ is a useful complement. 

First, we list some relations among the basic scalar harmonic $Y^A \equiv A$ and the basic vector harmonic $Y^A_{\a} \equiv A_{\a}$ on $S^d$:

\eqn\dua{\eqalign{\sum_{\a=1}^d A_{\a}B_{\a} &= -AB + \delta^{AB}\cr
\nabla_{\a}A &= A_{\a}\cr
\nabla_{\b}A_{\a} &= -\delta_{\a\b}A\cr}}

The index $A=1\ldots d+1$ runs over SO($d+1$), and $\a=1\ldots d$ over $S^d$ coordinates. The basic harmonics $A$ and $A_{\a}$ transform in the fundamental of SO($d+1$). All scalar harmonics $Y^{A_1\ldots A_k}$ are made from $A$, and all transverse vector harmonics $Y^{A_1\ldots A_{k+1}}_{\a}$ from $A$ and $A_{\a}$. These obey
\eqn\dub{\eqalign{\nabla^2Y^{A_1\ldots A_k} &= -k(k+d-1)Y^{A_1\ldots A_k}~, \quad k=1,2,\ldots\cr
\Delta_1 Y^{A_1\ldots A_{k+1}}_{\a} &= (k+1)(k+d-2)Y^{A_1\ldots A_{k+1}}_{\a}~, ~~ k=1,2,\ldots\cr}}
where  the scalar Laplacian $\nabla^2 = \nabla_{\a}\nabla^{\a}$, and we define $\Delta_p = d^{\dagger}\cdot d + d\cdot d^{\dagger}$ as the Laplace operator acting on transverse $p$-forms. 

Henceforth we switch to a notation $Y^k_p$ for transverse $p$-form harmonics of degree $k+p$ in the basic scalar harmonic: e.g. $Y^k_0 \equiv Y^{A_1\ldots A_k}$, and $Y^k_2 \equiv Y_{[\a\b]}^{A_1\ldots A_{k+2}}$.

We can write $\Delta_p$ in terms of the scalar Laplacian $\nabla^2$ as follows:
\eqn\duc{-\Delta_p = \nabla^2 + p(p-d)}
This is useful in confirming \dub\ using \dua.

Here we are concerned with scalars and transverse two-forms on $S^6$, whose spectra were provided in \dug\ and are recalled below,
\eqn\dugg{\eqalign{
\nabla^2 Y_0^\ell &= -\ell(\ell+5)Y_0^\ell ~, \quad \ell=1,2,\ldots\cr
\Delta_2Y_2^n &= (n+2)(n+3)Y_2^n~, \quad n=1,2,\dots\cr}} 
with eigenvalue degeneracies
\eqn\dug{\eqalign{D_0(\ell) &= {(2\ell+5)(\ell+4)!\over 5!\ell!}\cr
D_2(n) &={(2n+5)n(n+1)(n+4)(n+5)\over 12}\cr}}
which are just the dimensionalities of the SO(7) representations.
In terms of $\nabla^2$, we write the $p=2$ Laplace equation as
\eqn\duh{(\nabla^2-8)Y_2^n = -(n+2)(n+3)Y_2^n}
For the lowest harmonic $n=1$, 
\eqn\dui{\nabla^2Y_2^{n=1} = -4Y_2^{n=1}}

Now, our goal is to show that for a transverse two-form $\tau\sim Y_2$, the object
\eqn\duj{\Lambda(n) \equiv (n+2)^2|\tau|_{S^6}^2 + |d\tau|_{S^6}^2~, \quad n=1,2,\ldots}
does not source the scalar harmonic $Y_0^{l=2n}$. We will demonstrate this for $n=1$. 

Our first task is to find the explicit $n=1$ harmonic, which must be transverse and have the correct eigenvalue $\nabla^2=-4$. One can show by explicit computation, solely using \dua, that the following object satisfies these constraints:
\eqn\dul{\eqalign{Y_2^{n=1}=Y_{[\a\b]}^{ABC} &= A_{\a}B_{\b}C + B_{\a}C_{\b}A + C_{\a}A_{\b}B\cr&- A_{\b}B_{\a}C + B_{\b}C_{\a}A + C_{\b}A_{\a}B\cr
&= \Big(A_{\a}B_{\b}C + ({\rm Cyclic~ in~ A,B,C})\Big ) - (\a\leftrightarrow \b)\cr}}
To evaluate $\Lambda(n=1)$, we wish to write $\tau=Y_2^{n=1}$ and its totally antisymmetric field strength ${\cal{F}}=d\tau$ in components:
\eqn\dum{\tau = \tau_{\a\b}d\theta^{\a}\wedge d\theta^{\b}~, \quad {\cal{F}} = {\cal{F}}_{\gamma\a\b}d\theta^{\gamma}\wedge d\theta^{\a}\wedge d\theta^{\b}}
where $\theta^{\a}$ are $S^6$ coordinates, $d\theta^{\a}d\theta^{\a} = d\Omega^2_6$. Thus, we have
\eqn\dun{\eqalign{\tau_{\a\b} &= Y_{[\a\b]}^{ABC}\cr
{\cal{F}}_{\gamma\a\b} &= \nabla_{\gamma}\tau_{\a\b}+\nabla_{\a}\tau_{\b\gamma}+\nabla_{\b}\tau_{\gamma\a}\cr}}
We compute 
\eqn\duna{\eqalign{{\cal{F}}_{\gamma\a\b} = &3\Big[A_{\a}B_{\b}C_{\gamma} + B_{\a}C_{\b}A_{\gamma}  + C_{\a}A_{\b}B_{\gamma}\cr& - A_{\b}B_{\a}C_{\gamma}  + B_{\b}C_{\a}A_{\gamma}  + C_{\b}A_{\a}B_{\gamma}\Big]}}
Now we compute, for some fixed $(A,B,C)$:
\eqn\duo{\eqalign{|\tau|^2_{S^6} &= {1\over 2!}\tau_{\a\b}\tau^{\a\b} = A^2+B^2+C^2\cr
|d\tau|^2_{S^6} &= {1\over 3!}{\cal{F}}_{\gamma\a\b}{\cal{F}}^{\gamma\a\b} = 9(1-A^2-B^2-C^2)\cr}}
Therefore,
\eqn\dup{\Lambda(n=1) = 9|\tau|^2_{S^6}+|d\tau|^2_{S^6} = 9}
There are nontrivial cancellations in this result: given that $\tau_{\a\b}$ is of third degree in the basic scalar harmonic, one might have expected to source the $\ell=6$ scalar harmonic. 

As discussed in the main text, this result implies the existence of a family of four-dimensional Schr\"odinger solutions with $z=5/3$ and hyperscaling violation exponent $\theta=-1/3$. The constancy of $\Lambda(n=1)$
was perhaps to be expected from 
constructions of Sch$_5$ and Sch$_3$ solutions from null deformations of the D3 and D1-D5 systems, respectively \refs{\BobevMW, \DonosXC, \KrausPF, \BobevQX}. Those solutions required fluxes with vectors, rather than two-forms, on the internal manifold; nevertheless, when the vector was chosen to saturate its eigenvalue lower bound (i.e. to be Killing), $g_{++}$ had no dependence on the internal coordinates. All solutions obtained by TsT transformation of an AdS background have this property. As a side note, there is another solution in the literature with the same symmetries as our $n=1$ solution, also with $z=5/3$ and a constant $g_{++}$ \MazzucatoTR. It was obtained by TsT transformation of the relativistic D2-brane solution, and as such has $H$ flux instead of extra $F_4$ flux as we do here. Nevertheless they may be related by duality and dimensional reduction.

\listrefs\end